\documentstyle [12pt]{article}
\begin{document}
\setlength{\baselineskip}{.3in}
\setlength{\jot}{0.2in}
\title{Chromosome mapping: radiation hybrid data and stochastic spin models}

\begin{titlepage}

\vspace{1cm}

\begin{center}
{\large \bf Chromosome mapping: radiation hybrid data and stochastic spin models}
\end{center}

\vspace{1cm}

\begin{center}

{\bf C.T. Falk* and H. Falk**}\\

\vspace{1cm}

*The New York Blood Center, New York, NY; **Physics Dept., City College of
the City University of New York, New York, NY\\
\vspace{.5cm}
cathy@server.nybc.org\\
alfcc@cunyvm.cuny.edu\\
\end{center}

\vspace{3cm}

\begin{flushleft}

\vspace{3.25cm}

\vspace{4.5cm}
\end{flushleft}

\begin{flushleft}
\begin{center}
\section*{\hspace{-.5cm}Abstract\vspace{.5mm}}
\end{center}
This work approaches human chromosome mapping by developing      
algorithms for ordering markers associated with radiation hybrid data.
Motivated by recent work of Boehnke et al. [1], we formulate the
ordering problem by developing stochastic spin models to search for
minimum-break marker configurations. As a particular application, the
methods developed are applied to 14 human chromosome-21 markers
tested by Cox et al. [2]. The methods
generate configurations consistent with the best found by others.
Additionally, we find that the set of low-lying configurations is
described by a Markov-like ordering probability distribution. The
distribution displays cluster correlations reflecting closely linked 
loci. \\

\end{flushleft}


\end{titlepage}
\begin{center}
\section*{I. INTRODUCTION}
\end{center}
\setcounter{section}{1}
\renewcommand{\theequation}{\arabic{section}.\arabic{equation}}

     The use of data from radiation hybrid (RH) experiments has become
a useful method for fine structure mapping of human chromosomes.
Based on methods described by Goss and Harris [3, 4], Cox et al.
[2] and Burmeister et al. [5] have developed the technique
in detail so that results from their experiments provide material
for ordering DNA markers on human chromosomes. \\

     The basic strategy employed in radiation hybrid mapping (fully
described in Cox et al. [2]) entails
irradiating a rodent-human somatic cell hybrid, that contains a
particular human chromosome, with a lethal dose of x-rays.  This
will cause the chromosomes to break into several fragments.
After fusion with HPRT-deficient rodent cell lines, only the fused cells,
containing both the X-irradiated cells and the normal rodent
cells, will survive if grown in HAT medium. Each hybrid clone
arising from this fusion
will contain a unique set of fragments from the original human
chromosome.  Each clone can then be typed for a set of human DNA
markers (equivalently, loci) known to be on that human chromosome.  
Based on the
assumption that tightly linked markers are unlikely to be broken
apart by the radiation, markers close to one another will show a
correlated pattern of retention in the clones; whereas, distant
markers will be retained in a relatively independent manner. \\

     Several methods for ordering markers have been developed using
results from RH experiments [including both parametric (Cox et
al. [2]; Boehnke et al. [1]) and non-parametric methods 
( Boehnke et al. [1]; Falk [6]; Bishop and Crockford [7]; Weeks et al.
[8]). In particular, Boehnke et al. [1] used a mathematical
quantity associated with the number of breaks and then used
optimization techniques to minimize that quantity. 
One optimization technique involved a simulated annealing search for
configurations associated with minimal numbers of breaks. \\

 We set out to 
study and understand the work of Boehnke et al., and we
developed a formulation in the context of 
stochastic spin systems (Falk [9]).  It may be useful to point out
some differences in the formulations.\\

     Boehnke et al. use block inversions of a given
marker order and compare the old and new orders with respect to
"obligate" breaks. They then apply simulated annealing techniques
and decide, at each time step,  whether to retain the 
original order or transition to the new.  If a transition would result 
in a smaller number of breaks, the transition is made with probability one.  
If a transition would not decrease the number of breaks, then the 
transition probability is less than one, and that transition probability 
systematically decreases over time.\\

In our study we implement three algorithms which incorporate 
the number of "breaks". The three algorithms are three
stochastic spin models. These, too, are designed to search for 
configurations with
small numbers of breaks.  A probability is constructed to
determine whether or not to retain the current order or
transition to the new.  The probabilities are set up so
as to bias the decision towards transitions to configurations with fewer
breaks; however, at a given step, a possible transition leading
to a smaller number of breaks, will not necessarily be realized.\\

The spin language provides  mathematically intuitive
expressions for the number of breaks. Those expressions contain products 
of adjacent spin variables, and calculations involving breaks are easily 
presented in spin notation. For those seeking a rigorous mathematical 
setting, we remark that Liggett [10] has treated stochastic spin models
and related models from biology, physics and econonomics. Liggett's book 
contains an extensive bibliography and guides the reader to survey papers 
by Griffeath, Durrett, Stroock, Holley, and others.\\

\begin{center}
\section*{II. METHODS}
\end{center}
\setcounter{section}{2}
\renewcommand{\theequation}{\arabic{section}.\arabic{equation}}

     We are considering M clones, each of which is tested for the presence
(or absence) of N different DNA markers. It is convenient to represent
the clones as M rows, each with N sites. Thus, one pictures a two-dimensional 
array of M rows, N columns.\\

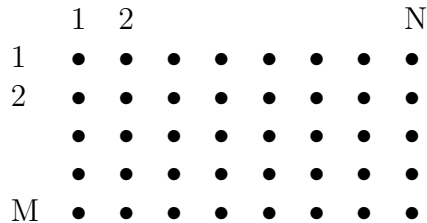
\vspace{1cm}
\begin{center}
\begin{tabular}{lllllllll}
    & 1     & 2     &       &       &       &       &       & N \\
   1 & $ \bullet $      & $ \bullet $      & $ \bullet $      & $ \bullet $      & $ \bullet $      & $ \bullet $      & $ \bullet $      & $ \bullet $  \\
  2 & $ \bullet $      & $ \bullet $      & $ \bullet $      & $ \bullet $      & $ \bullet $      & $ \bullet $      & $ \bullet $      & $ \bullet $  \\
    & $ \bullet $      & $ \bullet $      & $ \bullet $      & $ \bullet $      & $ \bullet $      & $ \bullet $      & $ \bullet $      & $ \bullet $  \\
    & $ \bullet $      & $ \bullet $      & $ \bullet $      & $ \bullet $      & $ \bullet $      & $ \bullet $      & $ \bullet $      & $ \bullet $  \\
  M & $ \bullet $      & $ \bullet $      & $ \bullet $      & $ \bullet $      & $ \bullet $      & $ \bullet $      & $ \bullet $      & $ \bullet $  
\end{tabular}
\end{center}
\vspace{1cm}

     Assign a particular DNA marker, labelled $f_j$, $(j = 1,2,...,N)$,
to each column. Associate a variable ("a spin") $s_{ij}$ with the site at
row $i$, column $j$. If the marker $f_j$ is present at site $(i,j)$, take
$s_{ij} = +1$; if $f_j$ is not present at site $(i,j)$, take $s_{ij} = -1$.
If one is uncertain as to whether a marker is present at site $(i,j)$, we take $s_{ij}$
to be unknown, and we deal with such sites in a manner to be specified 
subsequently.\\

     For $(j = 1,2,...,N)$ the marker $f_j$ was arbitrarily assigned to
column $j$. But any of the $N!$ assignments would be possible. A criterion
is needed for judging the "goodness" of an assignment.\\

     Considering that the DNA markers being tested for lie on a
particular human chromosome, those markers which are tightly
linked are likely to appear together or not appear in each
clone.  Therefore, it is reasonable to seek those marker assignments
which reflect such correlation. For that purpose we say that a "break"
exists between sites $(i,j)$ and $(i,j+1)$ whenever $s_{ij}s_{i,j+1} = -1$.
The strategy is to minimize the total number of breaks.
\nopagebreak
Note that in the
two-dimensional array of spins, a break refers only to horizontal,
nearest-neighbor spin pairs.
    
     Here we devise and test several algorithms which monitor the
total number of breaks while selectively permuting column labels. The
algorithms attempt to explore the vast configuration space in the manner 
of a "random walk", biased toward configurations having a reduced number 
of breaks. Notice,
for $N = 14$ there are $N! = 87178291200$ permutations of the numbers
$(1,2,...,14)$. Thus, in the spirit of the travelling salesman problem,
simulated annealing techniques are used (Kirkpatrick et al [11], Press, et al [12]). \\

{\bf Nearest-neighbor transposition algorithm}\vspace*{0.5cm}\\
 
\noindent(0) \hspace{0.25cm} Start with a specified configuration $\{s\}$ of the variables $s_{ij}$.\vspace*{0.5cm}\\
(1) \hspace{0.25cm} Select a number $j$ at random from the set $\{1,2,...,N\}$.\vspace*{0.5cm}\\
(2) \hspace{0.25cm} If $j \neq N$, compute the total number of 
breaks between columns $j-1, j$ and between columns $j+1, j+2$. Denote that
number by $B_j (1)$:\\

\begin{equation}
B_j(1) = \sum_{q=1}^M [(1-\delta_{j,1})\frac{1-s_{q,j-1}s_{qj}}{2}+
(1-\delta_{j,N-1})\frac{1-s_{q,j+1}s_{q,j+2}}{2}]
\end{equation}
\vspace{0.25cm}\\
\noindent for $j \in\{1,2,...,N-1\}$. The Kronecker delta contained in 
$(1-\delta_{i,j})$ is inserted to handle "end effects" since column
$1$ has no left neighbor and $N$ has no right neighbor.\vspace*{0.5cm}\\
(3) \hspace{0.25cm} Repeat (2) with the spin values $s_{ij}$ and $s_{i,j+1}$
interchanged for $i = 1,2,...,M$, and denote the resulting sum by $B_j(2)$
instead of $B_j(1)$.\vspace*{0.5cm}\\
(4) \hspace{0.25cm} Then compute $B_j$, the change in the number of breaks 
resulting from the interchange of the columns of spin values $s_{ij}$ 
and $s_{i,j+1}$.\vspace*{0.5cm}\\

\begin{equation}
B_j = B_j(2) - B_j(1) \mbox{ \hspace{1.25cm} for } j \in\{1,2,...,N-1\}.
\end{equation}

\noindent(5) \hspace{0.25cm} Interchange DNA marker column assignments 
and the associated spin values for 
columns $j$ and $j+1$ with probability.\vspace*{0.5cm}\\

\begin{eqnarray}
w_{1j} & = & \frac{\exp{(-\beta B_j/M)}}{2 \cosh{(\beta B_j/M)}}  \mbox{ \hspace {0.5cm} for } j \in\{1,2,...,N-1\} \nonumber \\ [0.8cm]
 & = & 0  \mbox{ \hspace{1.25cm} for } j = N. 
\end{eqnarray}
\vspace{0.25cm}\\
Do not interchange DNA marker column assignments and the associated spin values 
for columns $j$ and $j+1$ with
probability\\

\begin{equation}
w_{2j} = 1 - w_{1j}.
\end{equation}

Note:\\

\begin{eqnarray}
w_{2j} & = & \frac{\exp{(+\beta B_j/M)}}{2 \cosh{(\beta B_j/M)}}  \mbox{ \hspace {0.5cm} for } j \in\{1,2,...,N-1\} \nonumber \\ [1.0cm]
 & = & 1  \mbox{ \hspace{1.25cm} for } j = N. 
\end{eqnarray}
\vspace{0.25cm}\\

The "inverse temperature" parameter $\beta$ \hspace{.20cm} ( $ 0 \leq \beta < \infty $ )
is allowed to increase in an empirically determined manner as the algorithm is
implemented. Clearly very large $\beta$ strongly favors transitions 
reducing the number of breaks; whereas, a small, positive value of $\beta$  
makes transitions to increase or
to decrease the number of breaks almost equally likely. Why not just take
$\beta$ large at the outset? The answer is (Kirkpatrick et al [11], Press et al [12])
that setting $\beta$ large early in the calculation may cause the
algorithm to get "trapped" in a local minimum before performing a significant 
search of configuration space. \\
 
     Thus, the initially chosen value for $\beta$ and the manner of 
increasing that value constitute the "art" of simulated annealing.\vspace*{0.5cm}\\
(6) \hspace{0.25cm} Return to (1) and repeat the procedure starting with the
current configuration $\{s'\}$ of the variables $s_{ij}'$.
The procedure may be terminated after a specified number of iterations and/or
when the total number of breaks\\

\begin{equation}
B = \sum_{q=1}^M \sum_{j=1}^{N-1} \frac{1-s_{qj}s_{q,j+1}}{2}
\end{equation}
\vspace{0.25cm}\\
has realized  acceptably small values. For each run one retains the 
configurations associated with the smallest values of $B$. \\

     If $\beta$ were fixed, then the above procedure would define a 
finite-state, discrete-time Markov chain with transition probability
$p(\{s'\}\mid\{s\})$ from a configuration $\{s\}$ to configuration $\{s'\}$.
Explicitly\\

\begin{eqnarray}
p(\{ s'\} \mid \{ s\} ) &  =  & \frac{1}{N}\sum_{j=1}^N \left\{ \left[ \prod_{q=1}^M \left( \frac{1+s_{qj}'s_{q,j+1}}{2} \right)
\left( \frac{1+s_{q,j+1}'s_{qj}}{2} \right) \right.  \right. \nonumber \\ 
&  & \left. \times \prod_{k=1;k\neq j, j+1}^N \left( \frac{1+s_{qk}'s_{qk}}{2} \right) \right] w_{1j} \nonumber  \\
 &  & \left. + \left[  \prod_{q=1}^M \prod_{k=1}^N \left( \frac{1+s_{qk}'s_{qk}}{2} \right) \right] w_{2j} \right\}.
\end{eqnarray}
\vspace{0.5cm} \\

{\bf Two-column transposition algorithm}\vspace*{0.5cm}\\

A natural extension of the nearest-neighbor transposition algorithm involves
columns $k,j$ with $k > j+1$. Remove two numbers at random from the set
$\{1,2,...,N\}$. Denote the smaller number by $j$ and the 
larger by $k$. If $k = j + 1$, follow the previously described 
nearest-neighbor algorithm, starting with step (2). If $ k > j + 1$, 
proceed as follows:\\

Consider the total number of breaks between columns $j-1,j; j,j+1; \\
k-1,k; k,k+1$. Denote that number by $B_{jk}(1)$, where\\

\begin{eqnarray}
B_{jk}(1) & = &  \sum_{q=1}^M \left[ (1-\delta_{j,1})\frac{1-s_{q,j-1}s_{qj}}{2} + \frac{1-s_{qj}s_{q,j+1}}{2} \right. \nonumber \\
 & & \left. + \ \frac{1-s_{q,k-1}s_{qk}}{2} + (1-\delta_{k,N})\frac{1-s_{qk}s_{q,k+1}}{2} \right]. 
\end{eqnarray}
\vspace{0.5cm} \\

After interchanging spin values $s_{ij}$ and $s_{ik}$ for $i = 1,2,....,M$,
the number of breaks is \\

\begin{eqnarray}
B_{jk}(2) & = &  \sum_{q=1}^M \left[ (1-\delta_{j,1})\frac{1-s_{q,j-1}s_{qk}}{2} + \frac{1-s_{qk}s_{q,j+1}}{2}\right. \nonumber \\
 & & \left. + \ \frac{1-s_{q,k-1}s_{qj}}{2} + (1-\delta_{k,N})\frac{1-s_{qj}s_{q,k+1}}{2} \right]. 
\end{eqnarray}
\vspace{0.5cm} \\

Then the change (in the number of breaks) resulting from the interchange is
denoted by $B_{jk}$, where\\

\begin{equation}
B_{jk} = B_{jk}(2) - B_{jk}(1).
\end{equation}
\\

Now interchange DNA marker column assignments and the associated spin values
for columns $j,k$  for $k > j+1$
with probability \\

\begin{eqnarray}
w_{1jk} &  = & \frac{\exp{(-\beta B_{jk}/M)}}{2 \cosh{(\beta B_{jk}/M)}} \mbox{ \hspace {0.5cm} for } j \in\{1,2,...,N-2\}.
\end{eqnarray}
\\

Do not interchange DNA marker column assignments and the associated spin values for columns $j$ and $k$ with
probability \\

\begin{equation}
w_{2jk} = 1 - w_{1jk}.
\end{equation}
\vspace{0.5cm} \\

{\bf Block-flip algorithm}\vspace*{0.5cm}\\

As in the preceding algorithm, remove two numbers at random from the set
$\{1,2,...,N\}$. Denote the smaller number by $j$ and the 
larger by $k$. If $k = j + 1$, follow the previously described 
nearest-neighbor algorithm, starting with step (2). If $ k > j + 1$, 
proceed as follows:\\

The block consists of columns $j,...,k$. Before flipping the block, the
number of breaks involving columns $j-1,j$ and columns $k,k+1$ is
\samepage{
\begin{eqnarray}
B_{jk}^{\ block}(1) & = &  \sum_{q=1}^M \left[ (1-\delta_{j,1})\frac{1-s_{q,j-1}s_{qj}}{2} \right. \nonumber \\
 & & \left. + \ (1-\delta_{k,N})\frac{1-s_{q,k}s_{q,k+1}}{2} \right]. 
\end{eqnarray}
}
\vspace{0.5cm} \\

After flipping the block, the number of breaks is \\

\begin{eqnarray}
B_{jk}^{\ block}(2) & = &  \sum_{q=1}^M \left[ (1-\delta_{j,1})\frac{1-s_{q,j-1}s_{qk}}{2} \right. \nonumber \\
 & & \left. + \ (1-\delta_{k,N})\frac{1-s_{qj}s_{q,k+1}}{2} \right]. 
\end{eqnarray}
 \\
\vspace{0.5cm} \\

Then the change in the number of breaks is denoted by $B_{jk}^{\ block}$, where\\

\begin{equation}
B_{jk}^{\ block} = B_{jk}^{\ block}(2) - B_{jk}^{\ block}(1).
\end{equation}
\\

Now flip the block with probability\\

\begin{eqnarray}
w_{1jk}^{\ block} &  = & \frac{\exp{(-\beta B_{jk}^{\ block}/M)}}{2 \cosh{(\beta B_{jk}^{\ block}/M)}} \mbox{ \hspace {0.5cm} for } j \in\{1,2,...,N-2\}.
\end{eqnarray}
\\

Do not flip the block with probability\\

\begin{equation}
w_{2jk}^{\ block} = 1 - w_{1jk}^{\ block}.
\end{equation}
\vspace{0.5cm} \\

The preceding method is similar to the block inversion algorithm used
by Boehnke et al., but our transition probabilities differ from theirs.\\

{\bf Unknown site content}\vspace*{0.5cm}\\

In any clone one may have strings of one or more sites where the DNA 
marker associated with each site in a string is unknown. 
Thus, the spin values are unknown for the string. In the above algorithms
such unknowns are dealt with in the following way. \\

Consider the case where the left and right ends of a string connect
respectively to known spin $s_{left}$ and to known spin $s_{right}$. 
If $s_{left} =  s_{right}$, then all spins in the string are replaced by
 $s_{right}$. If $s_{left} \neq  s_{right}$, then a fair coin toss is
simulated. If the coin shows head (tail), then all spins in the string
are replace by $+1\ (-1)$. \\

If a string (of unknowns) contains an end spin, then replace all spins in the string
by the value of the connecting spin ($s_{left}$ or $s_{right}$), as
appropriate.\\

     With all spin values now specified, the number of breaks can be
calculated for any of the above algorithms, and the relevant 
transition probability can be evaluated. The simulated annealing
proceeds one step. After that step, all of the previously unknown
spin values are again regarded as unknown. [Note: Due to a possible
column interchange or block flip, those unknown spins, which previously 
belonged to  particular strings, may now belong to different strings.]
The above prescription for replacing unknown spin values by $+1$ or $-1$
is now repeated, and the transition probabilities are 
reevaluated. The simulated annealing proceeds another step, etc.
\begin{center}
\section*{III. APPLICATION}
\end{center} 
\setcounter{section}{3}
\setcounter{equation}{0}
\renewcommand{\theequation}{\arabic{section}.\arabic{equation}}

As an example, consider the data  presented by Cox et al. [2]
relating to 14 markers on chromosome 21 that were tested in 99 RH
clones.  The 14 markers are given in Table 1.  These are the same
data used by Boehnke et al. [1] and presented in detail in
their Table 1.  For our algorithms an entry of 1 in their table
becomes +1,  0 becomes  -1 and a "?" remains unknown and is
treated at each step as described above. \\

All three algorithms were applied to the data in the 99 $\times$ 14 matrix
for 200,000 iterations.  Initial values of $\beta$ and incremental
steps for increasing $\beta$ were chosen. This produced a set of
permutations with acceptably small values of B, the total number
of breaks for a particular configuration.  For each run, a ranked
set of marker permutations with the smallest values of B was
retained. \\

Table 2 lists the 2 distinct "best" orders found by each algorithm
in a representative run.  
As can be
seen, the first algorithm, where two nearest-neighbor columns are
transposed, does not result in permutations with values of B that
are as low as those reached by algorithms 2 and 3.  Although, in
principle, the nearest-neighbor transposition should allow for
visiting all possible permutations of the columns, in practice,
such exploration is not efficiently accomplished here. The large 
configuration space, and the 
empirical nature of selecting $\beta$
to implement simulated annealing provide a computational challenge
for the first algorithm.  The other
two algorithms display improved ability to achieve low B
values with the chosen set of parameters.  Additionally these
algorithms reach the same optimal order as that attained by
Boehnke et al. (see their Table 2), with the same number of
breaks.  The only difference is that we have retained all 14 markers;
whereas they combined markers \underline{S12} and \underline{S111}, since 
the latter markers were indistinguishable in the data matrix.  
Hence for each marker order in their Table 2 we would have 2 orders.\\

Although algorithms such as these do not assure that the marker
order with the smallest number of breaks is the correct order,
inspection of a set of low-lying states leads to some useful
information about the stability of clusters of markers that retain
their local ordering.  For example, consider the set of unique
permutations representing the 24 "best" orders obtained from a
series of runs of the three algorithms (Table 3).  We see, e.g.,
that \underline{S47} and \underline{SOD} are nearest neighbors in all 24 permutations and
appear as the last two markers in 16 permutations.  Similarly, the
triplet \underline{S46}-\underline{S4}-\underline{S52} is preserved in 21 permutations and of these,
positions 2,3 and 4 contain \underline{S46}, \underline{S4} and \underline{S52} respectively 15 times.
Based on observations such as these, we looked for a general
ordering property associated with sets of low-lying configurations.\\

{\bf Markovian-like property of low-lying configurations}\vspace*{0.5cm}\\

Let the DNA marker at site $j$ be denoted by $f_j$, where $f_j$ is
a member of the set \{\underline{S16}, \underline{S48}, \underline{S46}, 
\underline{S4}, \underline{S52},...\}. A configuration of sites is denoted 
by the ordered N-tuple $(f_1, f_2, ..., f_N)$. We have used underlining 
to distinguish a DNA marker such as \underline{S11} from a spin 
variable such as $s_{11}$. \\

For any configuration of sites, one can define strings $(f_i, f_{i+1}, ...$
$f_{i+m})$ with $i = 1,...,N$; $0 \leq m \leq{N-i}$. \\

Consider a collection $\cal C$ of distinct configurations. In the 
collection the probability of a string, $(f_i, f_{i+1}, ..., f_{i+m})$ 
is denoted by $P_{i,...,i+m}(f_i, f_{i+1}, ..., f_{i+m})$. \\

Now consider the Markovian-like approximation \vspace{0.5cm} \\
$P_{i,...,i+m}(f_i, f_{i+1}, ..., f_{i+m})  \approx  P_{i,i+1,i+2}(f_i \mid  
f_{i+1}, f_{i+2})P_{i+1,i+2,i+3}(f_{i+1} \mid f_{i+2}, f_{i+3})...$ 
\vspace{0.3cm} \\
$... P_{i+m-2,i+m-1,i+m}(f_{i+m-2} \mid f_{i+m-1}, f_{i+m})
P_{i+m-1, i+m}(f_{i+m-1},f_{i+m})$ \hspace{0.75cm} $(\star)$ \\

\noindent for $i = 1, ..., N-2 \  $ ; $\ 2 \leq m \leq N - i$. \\

\begin{center}
{\bf Example 1.} \\
\end{center}

Consider the collection of 24 distinct low-lying configurations given in 
Table 3. Let 
(the number of markers)
$N = 14$. Look at the cluster $(f_6 = F, f_7 = G, f_8 = H, f_9 = I, f_{10} = J)$
where $F$ denotes the DNA marker \underline{S11}, $G$ denotes 
\underline{S1}, H denotes \underline{S18}, $I$ denotes \underline{S8}, and
J denotes \underline{APP}. This marker assignment corresponds to the ordering
of the first row of Table 3. \\ 

For the above collection we find the frequency \\

\begin{eqnarray}
P_{6,7,8,9,10}(F,G,H,I,J) & = & 7/24
\end{eqnarray}
\\
and we also find the  frequencies \\

\begin{eqnarray}
P_{6,7,8}(F \mid G,H) & =  & 11/11 \\ 
P_{7,8,9}(G \mid H,I) & =  & 7/11 \\
P_{8,9,10}(H \mid I,J) & = & 11/11  \\
P_{9,10}(I,J)         & = &  11/24,
\end{eqnarray}
\\

\noindent so the  approximation $(\star)$ with $i = 6, m = 4$ is satisfied by the
observed frequencies.\\

    Similarly, for the above configurations 
$P_{7,8,9,10}(G,H,I,J)  =  7/24$ and \\

\begin{eqnarray}
P_{7,8,9,10}(G,H,I,J) & \approx  & P_{7,8,9,10}(G \mid H,I) P_{8,9,10}(H \mid I,J)
P_{9,10}(I,J) \nonumber  \\
& = & (7/11)(11/11)(11/24) \nonumber \\
& = & 7/24.      
\end{eqnarray}
\\

\begin{center}
{\bf Example 2.} \\
\end{center}

Consider the same collection of 24 distinct low-lying configurations
used in example 1. Look at the cluster $(f_2 = B, f_3 = C, f_4 = D, f_5 = E)$
where $B$ denotes the DNA marker \underline{S48}, $C$ denotes 
\underline{S46}, D denotes \underline{S4}, $E$ denotes \underline{S52}. \\

For the above collection we find the frequency \\

\begin{eqnarray}
P_{2,3,4,5}(B,C,D,E) & = & 12/24,
\end{eqnarray}
\\
and we also find the  frequencies \\

\begin{eqnarray}
P_{2,3,4}(B \mid C,D) & =  & 12/15 \\ 
P_{3,4,5}(C \mid D,E) & =  & 15/16 \\
P_{4,5}(D,E)         & = &  16/24,
\end{eqnarray}
\\

\noindent again, the approximation $(\star)$ is satisfied by the
observed frequencies.\\

    However, since

\begin{eqnarray}
P_{2,3}(B \mid C) & =  & 13/17 \\
P_{3,4}(C \mid D) & =  & 15/18 \\
P_{4,5}(D \mid E) & =  & 16/16   \\
P_{2,3,4,5}(B,C,D,E) & = & 12/24,
\end{eqnarray}
\\
the frequencies do {\it not} satisfy the standard Markovian approximation \\

\begin{eqnarray}
P_{2,3,4,5}(B,C,D,E) & \neq & P_{2,3}(B \mid C)P_{3,4}(C \mid D)
P_{4,5}(D \mid E)P_5(E) \nonumber \\
12/24 & \neq & (13/17)(15/18)(16/16)(16/24) \nonumber \\
0.50 & \neq & 0.425
\end{eqnarray}
\\
     A referee has kindly pointed out that in the context of Markov random
fields (Spitzer [13]) one could perhaps find a rigorous basis for the
observed Markov-like property. We thank the referee for this stimulating
suggestion.\\
\begin{center}
\section*{IV. DISCUSSION}
\end{center}
     The implementation of experimental techniques using RH data
provides a bridge between chromosome mapping data generated from
families and physical mapping data.  RH experiments allow for the
relative ordering of genetic markers that are too closely spaced
to be detected by family linkage analysis.  Additionally, it is
not necessary to have polymorphic markers in order to generate
useful information.  Although not providing the level of
resolution of physical mapping, RH mapping can complement and
confirm results generated by pulse field gel electrophoresis.\\

     Boehnke et al. [1] have presented a full discussion of the
advantages and disadvantages of parametric and non-parametric
ordering algorithms.  As they point out, algorithms that search
for minimum break configurations do not allow for estimates of
distances between markers, nor do they provide relative likelihoods
for one marker order over another.  However, as the present work
and the work of Boehnke et al. demonstrate, retention and
inspection of sets of low-lying configurations yield important insight
relating to the correlations of markers.\\

     In our study we were interested in learning what properties might
be present in a set of configurations with relatively few breaks.
It became obvious that the set of low-lying configurations showed
the clustering of particular markers.  That was reassuring, since
in complex optimization problems of the travelling salesman
variety, one typically ends up with a set of low-lying
configurations and never discovers the true absolute minimum.\\

One striking feature which we discovered here is that the set of
low-lying configurations is described by a Markov-like
probability distribution.  That distribution contains all of the
observed clustering of the markers.  If one enlarges the set
of low-lying configurations to include configurations with 
larger and larger numbers
of breaks, the validity of the Markov-like approximation
deteriorates.  We are not in a position to judge whether the relation
$(\star)$ is necessarily a deep or broad result, but we regard it as
interesting.\\
\begin{center}
\section*{ACKNOWLEDGMENTS}
\end{center}
\begin{flushleft}                               
\hspace{-.5cm}\rule{14.25cm}{.5mm}\vspace{1mm}\\

We would like to thank Lynh Wu for her programming help in this project.
This research was supported in part by Grant GM29177 from the National
Institutes of Health (CTF).\\
\end{flushleft}

\begin{flushleft}
\section*{\hspace{-.5cm}References\vspace{.5mm}}
\hspace{-.5cm}\rule{14.25cm}{.5mm}\vspace{1mm}\\

\hspace{-.5cm}[1 ] M. Boehnke, K. Lange, D.R. Cox,   
Am. J.Hum. Genet. {\bf 49}, 1174 (1991).\vspace{.18cm}\\

\hspace{-.5cm}[2 ] D.R. Cox, M. Burmeister, E.R. Price, S. Kim and  R.M. Myers,
Science {\bf 250}, 245  (1990).\vspace{.18cm}\\

\hspace{-.5cm}[3 ] S.J. Goss and H. Harris,  
Nature {\bf 255}, 680 (1975).\vspace{.18cm}\\

\hspace{-.5cm}[4 ] S.J. Goss and H. Harris, 
J. Cell Sci.  {\bf 25}, 39 (1977).\vspace{.18cm}\\

\hspace{-.5cm}[5 ] M. Burmeister, S. Kim, E.R. Price, T. de Lange, U. Tantravahi,
R.M. Myers, and  D.R. Cox, 
Genomics {\bf 9}, 19 (1990).\vspace{.18cm}\\

\hspace{-.5cm}[6 ] C.T. Falk, 
Genomics {\bf 9}, 120 (1991).\vspace{.18cm}\\

\hspace{-.5cm}[7 ] D.T. Bishop, G.P. Crockford, 
In: Genetic Analysis Workshop 7: Issues in Gene Mapping and
Detection of Major Genes.
J.W. MacCluer, A. Chakravarti, D.R. Cox,
D.T. Bishop, S.J. Bale, M.H. Skolnick (eds.), Cytogenet Cell Genet
{\bf 59}, 93 (1992).\vspace{.18cm}\\

\hspace{-.5cm}[8 ] D.E. Weeks, T. Lehner, J. Ott,
In: Genetic Analysis Workshop 7: Issues in Gene Mapping and
Detection of Major Genes.
J.W. MacCluer, A. Chakravarti, D.R. Cox,
D.T. Bishop, S.J. Bale, M.H. Skolnick (eds.), Cytogenet Cell Genet
{\bf 59}, 125 (1992).\\
\hspace{-.5cm} [9 ] H. Falk,
Physica, {\bf 104A}, 459 (1980).\vspace{.18cm}\\

\hspace{-.5cm} [10] T.M. Liggett, {\it Interacting Particle Systems} (Springer-Verlag,
New York, 1985).\vspace{.18cm}\\

\hspace{-.5cm} [11] S. Kirkpatrick, C.D. Gelatt and M.P. Vecchi,
Science {\bf 220}, 671 (1983).\vspace{.18cm}\\
{\samepage
\hspace{-.5cm} [12] W.H. Press, B.P. Flannnery, S.A. Teukolsky, and  W.T. Vetterling,
{\it Numerical  Recipes:  The  Art  of  Scientific Computing} (Cambridge Univ. Press, London, 1986) pp. 326-334.\vspace{.18cm}\\
\nopagebreak
\hspace{-.5cm} [13] F. Spitzer,
{\it Random Fields and Interacting Particle Systems} (Mathematical
Association of America, 1971); Am. Math. Monthly, {\bf 78}, 142 (1971).}

\end{flushleft}


\         \vspace{5cm}\\

\begin{center}
\bf Table 1.\\
\end{center}
\begin{flushleft}
\vspace{.5cm}
\begin{center}
Fourteen Chromosome-21 Markers Used in the Examples.\\
\end{center}
\vspace{1cm}
\  \\
\ \\
\vspace{.25cm}
\begin{tabular}{lrrrrrrrrrrrrr}
\hspace{-1cm} APP &  S1 &  S4 &  S8 & S11 & S12 & S16 & S18 & S46 & S47 & S47 & S52 & S111 &  SOD\\
\end{tabular}
\vspace{.5cm}

\hspace{-1cm}Note: \ All numbered loci have a prefix of D21; APP = amyloid precursor; \\ 
\hspace{-1cm}SOD = superoxide dismutase.\\
\end{flushleft}


\         \vspace{3cm}\\

\begin{center}
\bf Table 2.\\
\end{center}
\begin{flushleft}
\vspace{.5cm}
\begin{center}
Minimum-Break Orders for Representative Simulations.\\
\end{center}
\vspace{1cm}
\hspace{-1cm}Algorithm       \hspace{6cm} Marker Order\\
\hspace{-.75cm} $\downarrow$ \hspace{.30cm}Breaks\\
\vspace{.25cm}
\begin{tabular}{lrrrrrrrrrrrrrrrr}

\hspace{-1cm}     1  &      205  &     APP  &   S8  &   S1  &  S11  &  S16  &   S4  &  S12  &  S18  &  S48  &  S46  &  S52 S111  &  S47  &  SOD \\[.2in] 
\hspace{-1cm}     1  &      207  &     APP  &   S8  &   S1  &  S11  &  S16  &   S4  &  S12  &  S18  &  S46  &  S48  &  S52 S111  &  S47  &  SOD \\[.2in] 
\hspace{-1cm}        &           &          &       &       &       &       &       &       &       &       &       &            &       &      \\[.2in]
\hspace{-1cm}     2  &      123  &     S16  &  S48  &  S46  &   S4  &  S52  &  S11  &   S1  &  S18  &   S8  &  APP  & S111  S12  &  S47  &  SOD \\[.2in] 
\hspace{-1cm}     2  &      123  &     S16  &  S48  &  S46  &   S4  &  S52  &  S11  &   S1  &  S18  &   S8  &  APP  &  S12 S111  &  S47  &  SOD \\[.2in] 
\hspace{-1cm}        &           &          &       &       &       &       &       &       &       &       &       &            &       &      \\[.2in] 
\hspace{-1cm}     3  &      123  &     S16  &  S48  &  S46  &   S4  &  S52  &  S11  &   S1  &  S18  &   S8  &  APP  & S111  S12  &  S47  &  SOD \\[.2in] 
\hspace{-1cm}     3  &      123  &     S16  &  S48  &  S46  &   S4  &  S52  &  S11  &   S1  &  S18  &   S8  &  APP  &  S12 S111  &  S47  &  SOD \\ 

\end{tabular}
\vspace{.5cm}

\hspace{-1cm}See note for Table 1. \\[.2in]
\end{flushleft}

\begin{center}
\bf Table 3.\\
\end{center}
\begin{flushleft}
\hspace{.7cm}24 Unique Marker Orders With Relatively Small Numbers of Breaks\\
\hspace{-1cm}Breaks     \hspace{6cm} Marker Order\\
\vspace{.25cm}
\begin{tabular}{lrrrrrrrrrrrrrr}
\hspace{-1cm}123  &  S16 &   S48 &   S46 &    S4 &   S52 &   S11 &    S1 &   S18 &    S8 &   APP &  S111 &   S12 &   S47 &   SOD   \\[.1in] 
\hspace{-1cm}125  &  S16 &   S48 &   S46 &    S4 &   S52 &   S11 &    S1 &   S18 &   APP &    S8 &   S12 &  S111 &   S47 &   SOD   \\[.1in] 
\hspace{-1cm}126  &  S48 &   S16 &   S46 &    S4 &   S52 &   S11 &    S1 &   S18 &    S8 &   APP &   S12 &  S111 &   S47 &   SOD   \\[.1in] 
\hspace{-1cm}126  &  S48 &   S16 &   S46 &    S4 &   S52 &   S11 &    S1 &   S18 &    S8 &   APP &  S111 &   S12 &   S47 &   SOD   \\[.1in] 
\hspace{-1cm}127  &  S11 &    S1 &   S16 &   S48 &   S46 &    S4 &   S52 &   S18 &    S8 &   APP &  S111 &   S12 &   S47 &   SOD   \\[.1in] 
\hspace{-1cm}127  &  S16 &   S48 &   S46 &    S4 &   S52 &   S11 &    S1 &   S18 &    S8 &   APP &   S12 &  S111 &   S47 &   SOD   \\[.1in] 
\hspace{-1cm}127  &  S46 &   S48 &   S16 &    S4 &   S52 &   S11 &    S1 &   S18 &    S8 &   APP &   S12 &  S111 &   S47 &   SOD   \\[.1in] 
\hspace{-1cm}128  &  S11 &    S1 &   S52 &    S4 &   S46 &   S48 &   S16 &   S18 &    S8 &   APP &   S12 &  S111 &   S47 &   SOD   \\[.1in] 
\hspace{-1cm}128  &  S11 &    S1 &   S52 &    S4 &   S46 &   S48 &   S16 &   S18 &    S8 &   APP &  S111 &   S12 &   S47 &   SOD   \\[.1in] 
\hspace{-1cm}128  &  S16 &   S48 &   S46 &    S4 &   S52 &   SOD &   S47 &   S12 &  S111 &   APP &    S8 &   S18 &    S1 &   S11   \\[.1in] 
\hspace{-1cm}128  &  S16 &   S48 &   S46 &    S4 &   S52 &   SOD &   S47 &  S111 &   S12 &   APP &    S8 &   S18 &    S1 &   S11   \\[.1in] 
\hspace{-1cm}129  &  S11 &    S1 &   S16 &   S48 &   S46 &    S4 &   S52 &   S18 &    S8 &   APP &   S12 &  S111 &   S47 &   SOD   \\[.1in] 
\hspace{-1cm}129  &  S11 &    S1 &   S16 &   S48 &   S46 &    S4 &   S52 &   S18 &   APP &    S8 &  S111 &   S12 &   S47 &   SOD   \\[.1in] 
\hspace{-1cm}129  &  S16 &   S48 &   S46 &    S4 &   S52 &   S11 &    S1 &   S18 &   S12 &  S111 &    S8 &   APP &   S47 &   SOD   \\[.1in] 
\hspace{-1cm}129  &  S16 &   S48 &   S46 &    S4 &   S52 &   S11 &    S1 &   S18 &  S111 &   S12 &    S8 &   APP &   S47 &   SOD   \\[.1in] 
\hspace{-1cm}130  &  S16 &   S48 &   S46 &    S4 &   S52 &   APP &    S8 &   S12 &  S111 &   S47 &   SOD &   S18 &    S1 &   S11   \\[.1in] 
\hspace{-1cm}130  &  S16 &   S48 &   S46 &    S4 &   S52 &   APP &    S8 &  S111 &   S12 &   S47 &   SOD &   S18 &    S1 &   S11   \\[.1in] 
\hspace{-1cm}130  &  S16 &   S48 &   S46 &    S4 &   S52 &   SOD &   S47 &   APP &    S8 &   S12 &  S111 &   S18 &    S1 &   S11   \\[.1in] 
\hspace{-1cm}130  &  S16 &   S48 &   S46 &    S4 &   S52 &   SOD &   S47 &   S12 &  S111 &    S8 &   APP &   S18 &    S1 &   S11   \\[.1in] 
\hspace{-1cm}130  &  S16 &   S48 &   S46 &    S4 &   S52 &   SOD &   S47 &  S111 &   S12 &    S8 &   APP &   S18 &    S1 &   S11   \\[.1in] 
\hspace{-1cm}130  &  S16 &   S48 &   S46 &   S52 &    S4 &   SOD &   S47 &   S12 &  S111 &   APP &    S8 &   S18 &    S1 &   S11   \\[.1in] 
\hspace{-1cm}130  &  S48 &   S16 &   S46 &    S4 &   S52 &   S11 &    S1 &   S18 &   S12 &  S111 &   APP &    S8 &   S47 &   SOD   \\[.1in] 
\hspace{-1cm}130  &  S52 &    S4 &   S16 &   S48 &   S46 &   S11 &    S1 &   S18 &    S8 &   APP &   S12 &  S111 &   S47 &   SOD   \\[.1in] 
\hspace{-1cm}130  &  S52 &    S4 &   S46 &   S48 &   S16 &   S11 &    S1 &   S18 &    S8 &   APP &   S12 &  S111 &   S47 &   SOD   \\
\end{tabular}
\vspace{.5cm}

\hspace{-1cm}See note for Table 1. \\
\end{flushleft}

\end{document}